\documentclass[twocolumn,showpacs,groupedaddress,amsmath,amssymb]{revtex4-2}
\usepackage{graphicx,graphicx,epsfig}
\usepackage{epsfig}
\usepackage[usenames]{color}
\usepackage{bm}
\usepackage{enumerate}
\usepackage[T1]{fontenc}
\usepackage[colorlinks,
linkcolor=blue,
anchorcolor=blue,
citecolor=blue
]{hyperref}
\usepackage{textcomp}

\begin{document}
	
	\title{Field-driven side-by-side magnetic domain wall dynamics in ferromagnetic nanostrips}
	\author{Zhoujian Sun}
\altaffiliation[]{These authors contributed equally to this work.}
	\author{Panpan Fang}
\altaffiliation[]{These authors contributed equally to this work.}
	\author{Xinwei Shi}
	\author{X. S. Wang}
\email[Corresponding author: ]{justicewxs@hnu.edu.cn}
	\author{Fuxiang Li}
\email[Corresponding author: ]{fuxiangli@hnu.edu.cn}
    \affiliation{School of Physics and Electronics, Hunan University, Changsha 410082, China}
	\date{\today,\now}
	\begin{abstract}

There has been a plethora of studies on domain wall dynamics in magnetic nanostrips, mainly because of its versatile non-linear physics and potential applications in data storage devices. However, most of the studies focus on out-of-plane domain walls or in-plane head-to-head (tail-to-tail) domain walls. Here, we numerically study the field-driven dynamics of in-plane side-by-side domain walls in ferromagnetic strips, which can be stable in the presence of an in-plane easy-axis anisotropy transverse to the strip. The domain walls move in a rigid-body manner at low field, and show complex Walker breakdown behavior at high field. We observe a multi-step Walker breakdown through vortex nucleation in wide strips. In the presence of Dzyaloshinskii-Moriya interaction (DMI), the first Walker breakdown field first decreases then increases with interfacial DMI, while keeps increasing with bulk DMI. These findings complement the current understanding on domain wall dynamics.

\end{abstract}
	
\date{\today}
\maketitle
	
\section{\label{sec:level1}INTRODUCTION}
Over the past few decades, domain wall  dynamics has attracted much attention because of the fundamental interests and potential applications in future memory and logic devices \cite{DW1,DW2,DW3}. Domain walls are transition regions between two differently oriented magnetic domains. The most traditional way to drive the domain wall motion is to apply an external field along one of the domains \cite{Field1,Field2,Field3,Field4}. Later, spin-polarized electric current \cite{Current1,Current2,Current3,Current4}, temperature gradient \cite{TG1,TG2,TG3,TG4}, and spin waves \cite{SW1,SW2,SW3} are used as control knobs of domain wall dynamics.

Due to the non-linear nature of magnetization dynamics, although the field-driven domain wall dynamics has been studied for the longest time, the understanding is still incomplete. On one hand, most analytical models are one-dimensional (1D) models that are only applicable for very thin and narrow lines or strips, or very large bulk systems which are uniform in the other two dimensions \cite{Field1,Tatara04,Shibata2011}. For generic systems, especially magnetic strips whose width is much larger than the domain
wall width, the 1D model fails due to the variation in the width direction \cite{Yuan2014,Laurson20151,Laurson20152}.
On the other hand, most studies focus on in-plane head-to-head (or tail-to-tail) domain walls \cite{SW1,TG4} or perpendicular magnetic anisotropy (PMA) domain walls \cite{Miron2011, Emori2013, Laurson20191}. This is mainly because these two types of domain walls are energetically preferable in magnetic nanostrips, which are the platform of domain wall racetrack memory \cite{DW2}. For soft magnetic materials, the domains align along the strip due to shape anisotropy, forming head-to-head (or tail-to-tail, HtH/TtT for short) domain walls. For materials with strong PMA, the domains are oriented out-of-plane, and the domain walls between them are PMA domain walls.
However, there is another type of 180\textdegree
domain walls which is usually ignored, i.e. the ``side-by-side'' domain walls \cite{FE2020}, as schematically depicted in Fig. \ref{fig1}. The magnetization directions of the two domains are in-plane and perpendicular to the strip. This type of domain walls may
exist in wide strips made of materials with in-plane magnetocrystalline anisotropy, such as cobalt or permalloy grown on some
designed substrates \cite{Cobalt2000,Py2020}.
	
	There have been a lot of studies on the field-driven dynamics and internal structures of head-to-head (tail-to-tail) domain walls and PMA domain walls \cite{Yuan2014,Laurson20151,Laurson20152}. It is well known that in a biaxial system with an easy anisotropy axis and a hard anisotropy axis, the domain wall moves under the longitudinal magnetic field in a rigid-body manner and the velocity increases
with the field strength. Beyond a critical field, the rigid-body motion breaks down, and the velocity drops with increasing field. This
behavior is called ``Walker breakdown'' and the critical field is called ``Walker breakdown field''. For thin strips, the breakdown occurs
when the domain wall rotates around the external field \cite{Field1}. For wide and thick strips, the dynamics is much more complicated.
It has been shown that the HtH/TtT or PMA domain walls undergo periodic transformation through the generation and annihilation of vortices \cite{Yuan2014,Laurson20151} or Bloch lines \cite{Laurson20152,Ono2016}. In this paper, we theoretically investigate the
field-driven dynamics of the third type of domain walls, i.e. the in-plane side-by-side domain walls \cite{Filippov2004}. By micromagnetic
simulations, we find a normal Walker breakdown in thin strips which compares well with the 1D analytical model, and a multi-step Walker
breakdown behavior in wide strips. The multi-step breakdown occurs due to the generation of vortices and the generation and annihilation
of vortex-antivortex pairs. Furthermore, we study the influence of Dzyaloshinskii-Moriya interaction (DMI) on the domain wall dynamics.
For bulk DMI (BDMI), the Walker breakdown field increases with BDMI strength. For interfacial DMI (IDMI), the Walker breakdown field first decreases then increases with increasing IDMI strength. These behaviors are different from those in HtH/TtT domain walls
\cite{Zhuo2016} and PMA domain walls \cite{Thiaville2012} due to the different domain wall geometry. Our findings fill in the blank
of side-by-side domain wall dynamics and complement the understandings on 180\textdegree domain wall dynamics.

\section{MODEL AND STATIC SIDE-BY-SIDE DOMAIN WALL}

\begin{figure}[ht]
\centering
\includegraphics[width=0.45\textwidth]{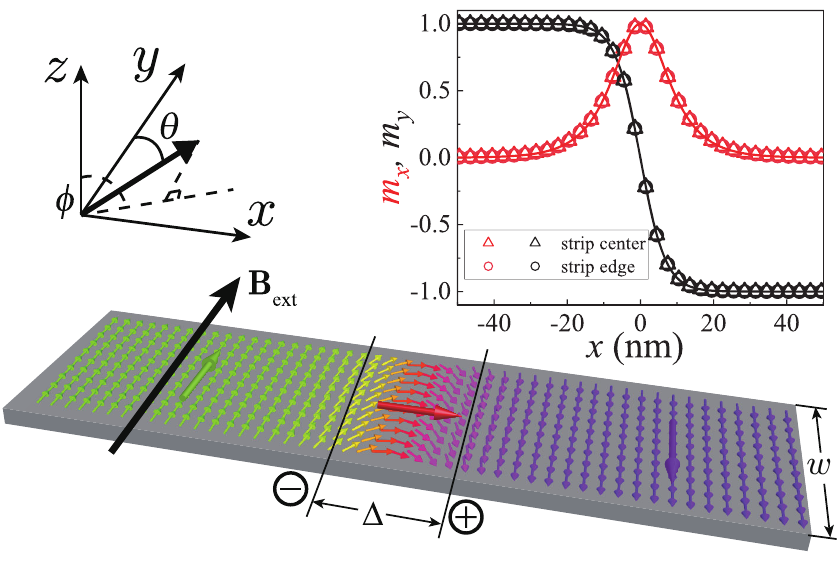}
\caption{Schematic illustration of an in-plane side-by-side magnetic domain wall. In the absence of DMI, the static domain wall is N\'{e}el-type in which all the spins are in-plane due to the shape anisotropy. The external field $\mathbf{B}$ is applied along $y$ direction. Inset: comparison of the $m_y$ and $m_x$ components at the edge and the centerline of the strip.}
\label{fig1}
\end{figure}

We consider a wide ferromagnetic strip of saturation magnetization $M_s$
along $x$ direction with easy axis along $y$ direction of strength $K_u$, as shown in Fig. \ref{fig1}. The length,
width and thickness of the strip are $l$, $w$ and $d$, respectively. To be convenient, we define a polar coordinate
system with respect to $y$ axis.

The magnetization dynamics is governed by the Landau-Lifshitz-Gilbert (LLG) equation \cite{LLGGilbert},
\begin{equation}
\frac{\partial \mathbf{m}}{\partial t}=-\gamma\mathbf{m}\times\mathbf{H}_{\mathrm{eff}} +\alpha\mathbf{m}\times\frac{\partial \mathbf{m}}{\partial t},
\label{LLGeq}
\end{equation}
where $\mathbf{m}$ is the unit vector of the magnetization direction, $\gamma$ is the gyromagetic ratio,
and $\alpha$ is the Gilbert damping.
The effective field $\mathbf{B}_\text{eff}$ is the variation of the total energy $E$, $\mathbf{B}_\text{eff}=
-\frac{\delta E}{M_s \delta \mathbf{m}}$. We do not consider the DMI at this stage. The total energy density $E$ is,
\begin{equation}
E=A|\nabla \mathbf{m}|^2-K_u m_y^2-M_sBm_y+E_d,
\label{totenergy}
\end{equation}
where $A\left|\nabla \mathbf{m}\right|^2$ is the exchange energy density with exchange constant $A$, $-K_u m_y^2$ is the easy-axis anisotropy energy density, $-M_sBm_y$ is the Zeeman energy density with $B$ the field strength along $y$ direction, and $E_d$ is the demagnetization energy density. As a demonstration of concept, we consider ferromagnet strips of thickness $d = 3$ nm with widths $w$ ranging from 48 to 2100 nm, and use a moving simulation window of length $l = 3072$ nm centered around the domain wall. The material parameters are $K_u = 2\times10^5\ \rm J/m^3$, $A =10^{-11}$ J/m, $\alpha=0.1$, $M_s=3\times10^5$ A/m.
The exchange length $L_\text{ex}=\sqrt{A/(\mu_0M_s^2)}\approx13.3$ nm. In our numerical simulations,
we use the Mumax3 \cite{mumax} package
to numerically solve the LLG equation \eqref{LLGeq} with the mesh size 3 nm $\times$ 3 nm $\times$ 3 nm, which is smaller than the
exchange length.

Due to the thin-film shape anisotropy, the domain wall center should be in-plane, resulting in a N\'{e}el wall. This can be easily
verified by numerical simulation. We set up an $\mathbf{m}\parallel +y$ ($-y$) domain at the left (right) half, and a region of $5L_\text{ex}$ wide with random magnetization in between the two domains as the initial state. After relaxing, the domain wall is
a N\'{e}el type one as shown in Fig. \ref{fig1}. For $w=48$ nm, the three components $m_{x,y,z}$ along the strip
edge ($y=-w/2$) and the centerline $y=0$) are plotted in the inset (the origin is set at the center of the strip).
It can be observed that the magnetization is almost uniform in the transverse ($y$) direction,
which is different from the strawberry-shape head-to-head walls \cite{Yuan2014} but similar to the PMA domain walls \cite{Emori2013}.
The domain wall center can either be $+x$ or $-x$ direction with a degenerated energy in the absence of DMI.
The domain wall profile can be well fitted by the Walker solution \cite{Field1},
\begin{equation}
\theta(x) = 2\arctan (e^{\frac{x-X}{\Delta}}),\quad \phi(x)=\frac{\pi}{2},
\end{equation}
where $X$ is the domain wall center position and $\Delta$ is the domain wall width.
Here, $X$ is set to 0 and $\Delta=6.86$ nm from the fitting. In Cartesian coordinates, the solution
is $m_y=-\tanh \frac{x-X}{\Delta}$, $m_x=\mathrm{sech} \frac{x-X}{\Delta}$, and $m_z=0$, as shown
by the solid lines in the inset of Fig. \ref{fig1}, showing good agreement with the numerical
data points.

The domain wall width $\Delta$ is related to $A$ and the effective easy-axis ($y$ axis) anisotropy $K_y$ by
$\Delta=\sqrt{\frac{A}{K_y}}$, where $K_y$ includes the magnetocrystalline anisotropy $K_u$ and the shape
anisotropy. The shape anisotropy is an approximation of the demagnetization effects which only considers the
homogeneous part of the demagnetization fields and ignores the inhomogenous part. The effective anistropy coefficients
along $y$ and $z$ axes are $-\frac{N_y-N_x}{2}\mu_0M_s^2$ and $-\frac{N_z-N_x}{2}\mu_0M_s^2$, respectively, where
$N_{x,y,z}$ are demagnetization factors related to the geometry \cite{demagfactor}. In a N\'{e}el wall configuration,
there are finite bulk magnetic charges $\rho=-M_s\nabla\cdot\mathbf{m}$ at the two sides of the domain wall as
schematically labeled in Fig. \ref{fig1}. Thus, the shape anisotropy is approximately that of a prism with dimensions $\Delta$, $w$,
and $d$ \cite{demagfactor,Mougin2007}. The total effective easy-axis anisotropy is $K_y=K_u-[N_y(\Delta,w,d)-N_x(\Delta,w,d)]$,
and the total effective hard-axis anisotropy is $K_z=-\frac{N_z-N_x}{2}\mu_0M_s^2$ which keeps the static domain wall in-plane.
The domain wall width $\Delta$ satisfies,
\begin{equation}
\Delta=\sqrt{\frac{A}{K_u-[N_y(\Delta,w,d)-N_x(\Delta,w,d)]\mu_0M_s^2/2}},
\end{equation}
where $\Delta$ can be self-consistently solved. With our parameters, we have $\Delta=6.82$ nm, which is quite closed to the
numerically fitted value 6.86 nm. The demagnetization factor $N_x-N_y$ increases with the strip width $w$.
Thus, $K_\text{eff}$ becomes larger and $\Delta$ becomes smaller for wider strips, although the change is insignificant
because $K_u$ dominates the total $K_\text{eff}$.

\section{\label{MTWB}MAGNETIC FIELD DRIVEN DOMAIN WALL DYNAMICS AND MULTI-STEP WALKER BREAKDOWN}

\begin{figure}[ht]
	\includegraphics[width=0.45\textwidth]{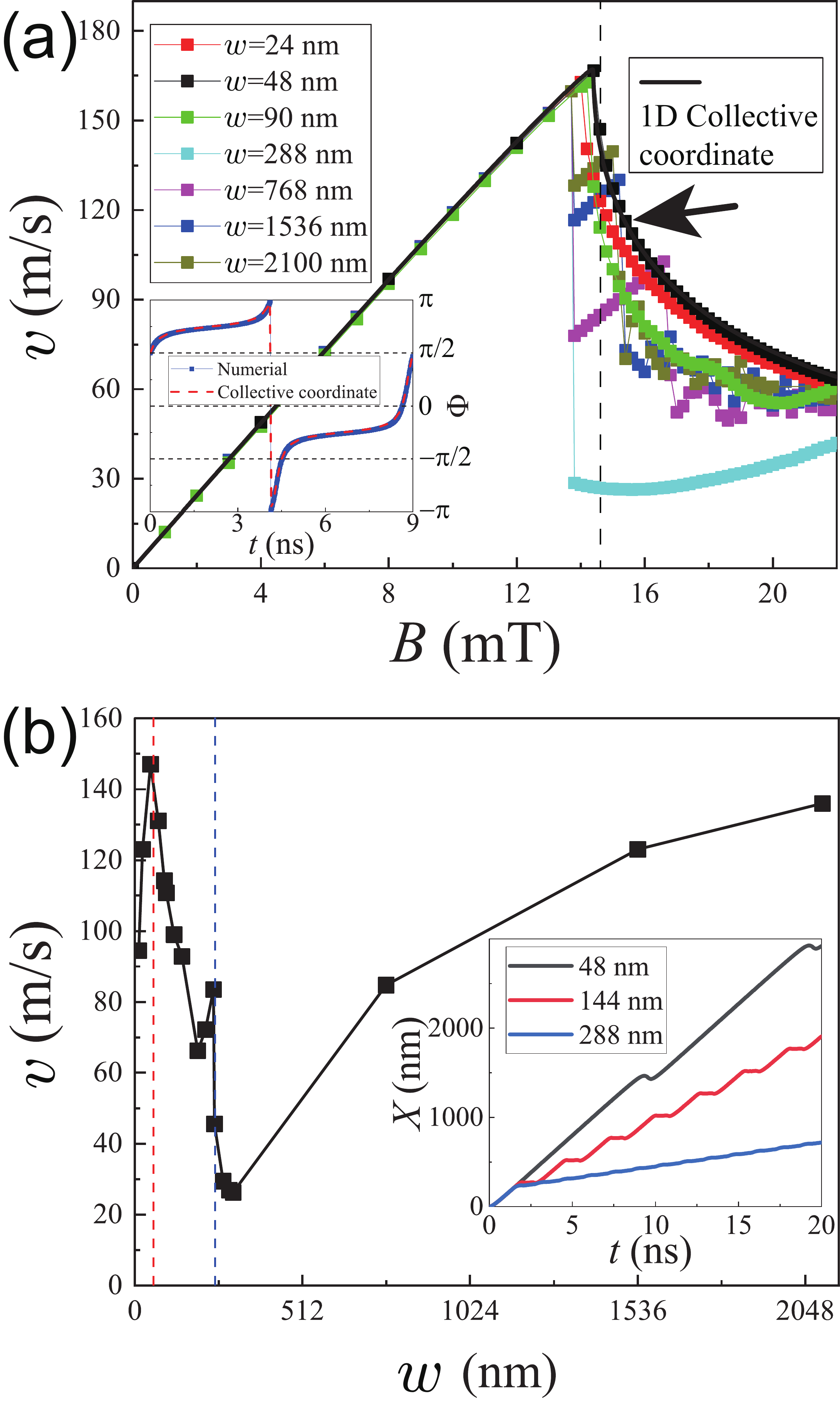}
	\caption{Domain wall velocity $v$ versus applied field $B$ for different strip width $w$. The black solid line is the result of
collective coordinate model with fitting parameters $K_y=2.09\times10^5$ J/m$^3$ and $K_z=4.337\times 10^4$ J/m$^3$ for $w=48$ nm.
The inset shows the time evolution of the azimuthal angel $\Phi$ of the domain wall plane for $B=15$ mT. The numerical data
(thick blue symbols) compares well with the collective coordinate model results (red dashed line).}
	\label{fig2}
\end{figure}

We then apply an external field $\mathbf{B}$ along $y$ direction to investigate the field-driven dynamics of the side-by-side wall.
Figure. \ref{fig2}(a) shows the domain wall speed $v$ versus the applied field for different strip width $w$. For thin strips ($w=24$ nm
and 48 nm in the figure), the speed shows a Walker-like behavior \cite{Field1}: below the Walker breakdown field, the domain wall propagates
like a rigid-body; above the Walker breakdown field, the domain wall rotates and oscillates. The average speed in the oscillation regime
can be obtained using a well-known 1D collective-coordinate model (CCM) \cite{Shibata2011,Thiaville2006}.
Because of the non-uniform out-of-plane component, the shape anisotropy is no longer a satisfactory approximation. By fitting the $w=48$
nm curve with the Walker formula and the collective coordinate results \cite{Thiaville2006}, we can obtain effective easy-axis anisotropy
$K_y=2.09\times10^5$ J/m$^3$ and hard-axis anisotropy $K_z=4.337\times 10^4$ J/m$^3$, and the fitted curve shows good agreement with the numerical data. The inset shows the time evolution of the average domain wall azimuthal angle $\Phi=\arctan {\frac{\langle m_x \rangle}{\langle m_z\rangle}}$ for $B=15$ mT (marked by the arrow in the main figure). The good agreement with the 1D collective coordinate model means that the side-by-side domain wall dynamics is still quasi-1D at strip width $w=48$ nm.

When the strip width gets wider, the dynamics in the transverse direction becomes more and more inhomogeneous. The way that domain wall
chirality periodically flips changes gradually from coherent rotation to vortex generation, and the velocity drop after the Walker
breakdown becomes sharper at the same time, which is similar to the well-studied PMA domain walls and HtH/TtT domain walls \cite{Yuan2014,Laurson20152,Laurson20191}. The reason is that the domain wall propagation velocity is proportional to the dissipation rate
of the Zeeman energy $E_\text{Zee}$, $\frac{d E_\text{Zee}}{dt}\propto M_sBv$ \cite{WXR2009}. When the domain wall moves rigidly, the Zeeman energy dissipation is the only way of energy dissipation. Beyond the Walker breakdown, when the internal dynamics of the domain wall becomes more complex, the Zeeman energy can be temporarily stored in the tilting or deformation of the domain wall. So the average Zeeman energy dissipation rate becomes lower,
resulting in a drop in velocity. However, different from the PMA domain walls \cite{Laurson20191}, the
Walker breakdown field does not change much. That is because although the dynamics is not quasi-1D, the Walker breakdown field
$B_w$ is still close to the 1D-model value $B_w=\frac{\alpha K_z}{M_s}$. In the side-by-side configuration, the hard axis is dominated
by the shape anisotropy $-\frac{N_z(\Delta,w,d)-N_x(\Delta,w,d)}{2}\mu_0M_s^2$. The change of $w$ in $y$ direction does not affect this
value too much.

	\begin{figure*}[!htp]
		\includegraphics[width=0.96\textwidth]{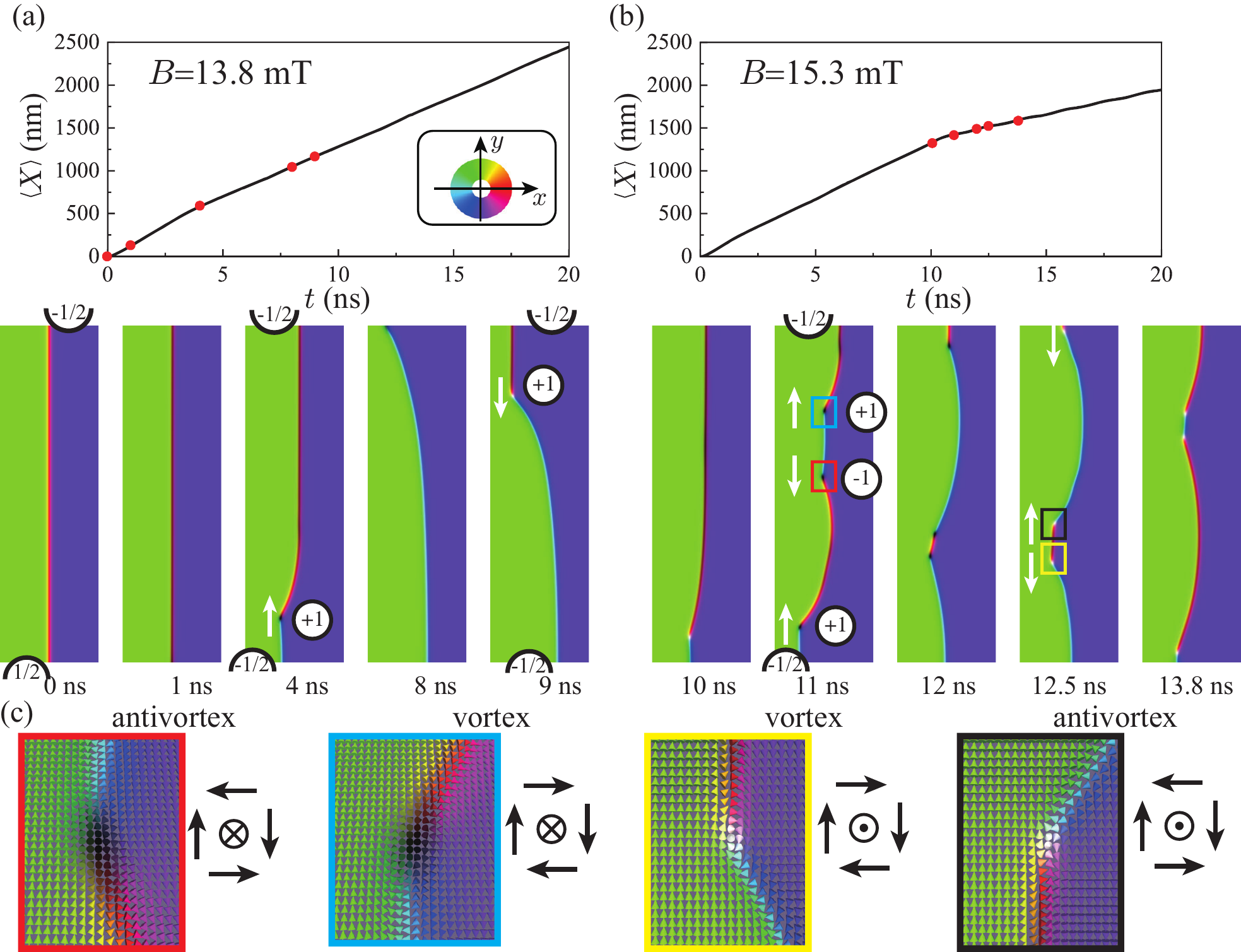}
		\caption{Upper panel: The average domain wall position $\langle X\rangle$ against time for $w=1536$ nm and (a) $B=13.8$ mT, (b)
$B=15.3$ mT. Lower panel: The snapshots of the magnetization texture near the domain wall corresponding to the moments marked by red dots in
the upper panel. The in-plane angle of the magnetization is encoded in the color ring shown in the inset. The winding numbers are labelled by circles or semicircles for vortices and edge defects. The moving directions of the vortices are indicated by arrows. Movies for the domain wall dynamics are shown in the Supplemental Material \cite{SM}. (c) Close-up textures of the (anti)vortices labelled by
frames of different colors in (b). The corresponding schematic spin configuration is schematically illustrated at the right hand side
of each texture.}
		\label{fig3}
	\end{figure*}
	
We can find more interesting and sophisticated behaviors by observing the details of the dynamics.  {Figure \ref{fig2}(b) shows the domain wall speed at $B=14.6$ mT (just beyond the Walker breakdown field) for different strip width $w$. There are roughly three regions,
separated by the red and blue dashed lines. For strips no wider than 48 nm (at the left hand side of the red dashed line), the 1D CCM is approximately valid. Thus, according to the 1D CCM and the effective anisotropy discussed above, the wider the strip, the
larger the Walker breakdown field. So the DW velocity increases with the width for a fixed field just above the breakdown field. No vortex is
generated and the DW motion is periodic due to the almost synchronized rotation of DW center.  The inset of Fig. \ref{fig2}(b) shows how the average domain wall position $\langle X\rangle$ calculated from $\langle X\rangle=\langle m_x\rangle L/2$ \cite{DW2004,mumax} moves with time. The black line is for $w=48$ nm. In each period, there is a long fast-moving
stage in which the velocity equals to the maximum rigid-body velocity at the Walker breakdown. At this stage, the domain wall almost keeps a
rigid-body motion with $\Phi\approx \pi/4$, where $\Phi$ is the azimuthal angle of the domain wall plane (see Appendix \ref{AA}). There is
also a short slow-down stage due to the rotation of the domain
wall center. $\Phi$ quickly rotates from $\Phi\approx \pi/4$ to $\Phi\approx -\pi/4$ at this stage and the domain wall moves fast again in a new period.}

 {For widths between the red dashed line and blue dashed line, there will be clear vortex generation and propagation, which is
similar to that observed in HtH/TtT  \cite{Yuan2014,Laurson20152} and PMA domain walls \cite{Laurson20191}. As shown in the inset of Fig. \ref{fig2}(b) by the red line, starting from a N\'{e}el wall pointing to $+x$ as shown in Fig. \ref{fig1}, the domain wall still
tilts out of plane, shrinks, accelerates, and moves fast with an almost uniform $\Phi$ angle at the beginning. But soon, a vortex whose center points to $-z$ (polarity $-1$) appears at the bottom edge, and the domain wall decelerates. The reason why the vortex appears at the bottom edge is that for the dipolar fields stabilize (destabilize) the $+x$ domain wall center at the top (bottom) edge. So for the domain wall center point to $-x$, the vortex is generated at the top edge. The Zeeman energy converts to the energy of the vortex and at the same time, the longitudinal motion slows down and wanders around. After the vortex annihilates at the other side, the domain wall moves fast again in the absence of vortex. After a while, a vortex (of opposite polarity) is generated at the top edge, and the motion slows down again.
For widths larger than the value indicated by the blue dashed line, there is also vortex generation and the starting stage is similar to the previous case. However, after the vortex annihilates, a new vortex \textit{immediately} appears at almost the same place. There is no fast-moving stage. Thus, the average speed is significantly reduced, as shown in the inset of Fig. \ref{fig2}(b) by the blue line for $w=288$ nm.} This behavior can also be observed in wider domain walls, but the domain wall
acceleration and deceleration becomes insignificant. This is because the energy cost of the vortex generation does not scale with the width since it is a local process, while the Zeeman energy is proportional to the width. So the impact of vortex generation on the Zeeman energy changing rate becomes weaker. Figure \ref{fig3}(a) shows the magnetization snapshots of a $w=1536$ nm strip under $B=13.8$ mT together with the average domain wall position $\langle X\rangle$ calculated from $\langle X\rangle=\langle m_x\rangle L/2$ \cite{DW2004,mumax}. The above mentioned single vortex generation (annihilation) processes associated with polarity flipping
at the strip edges can be clearly seen. For clarity of narration, we call such processes ``single-vortex processes''.

For  {strips wider than 300 nm}, a multi-step Walker breakdown can be observed, similar to that in PMA strips \cite{Laurson20191} [see data for $w=768$, 1536, and 2100 nm in Fig. \ref{fig2}(a)]. When further increasing the applied field after the first breakdown, a second breakdown occurs with a velocity drop.
To see what happens, we plot the magnetization snapshots of the 1536 nm wide strip under $B=15.3$ mT, just beyond the second breakdown
in Fig. \ref{fig3}(b), and show several typical snapshots of the magnetic texture near the domain wall.
During the first $\sim 10$ ns, the domain wall still undergo a single-vortex process as depicted above (see the 10 ns snapshot). However, after 10 ns, a vortex-antivortex pair of polarity $-1$  emerges inside the domain wall (see the 11 ns snapshot). The vortex (antivortex) has a winding number of $+1$ ($-1$) \cite{Oleg2005}. The winding number of a vortex is also called ``vorticity''. Due to the opposite vorticity and same polarity, the vortex and the antivortex have opposite gyrovectors so they move along $+y$ and $-y$ respectively according to the Thiele equation \cite{Thiele,Yuan2015}.
Also due to the finite gyrovector, the longitudinal speed of the vortex (antivortex) is smaller than the transverse domain wall, so the domain wall speed is slowed down. Then the vortex hits the top edge, reverses its polarity and moves down.
The antivortex annihilates with the other vortex coming up from the bottom edge, and a vortex-antivortex pair of polarity
$+1$ emerges at the same place (see the 12 ns and 12.5 ns snapshots). The vortex (antivortex) moves along $-y$ ($+y$), and then hits the bottom edge (annihilate with another
vortex), finishing a period. The vortex-antivortex generation and annihilation occur in the interior of the domain wall, and we call them ``two-vortex processes''. Figure \ref{fig3}(c) shows the zoom-in textures near the (anti)vortices labelled by frames of different colors in (b). To see the vorticity and the polarity more clearly, the spin configuration of each texture is schematically illustrated. The red, blue, yellow, black frames enclose antivortex of polarity $-1$, vortex of polarity $-1$, vortex of polarity $+1$, antivortex of polarity $+1$, respectively. Notice that the two-vortex processes in the interior and the single-vortex processes at the edges are
asynchronous. So after time goes on, the vortex-antivortex pair may appear at different positions inside the domain wall and
the dynamics may become more and more complicated.
We also label the winding numbers of all the vortices ($+1$ for vortices and $-1$ for antivortices) and edge defects. No matter how complicated the domain wall transformation is, the total winding number remains zero. When $B$ further increases, there can be more pairs of vortices and antivortices.
For larger applied field, more two-vortex processes occur at the same time, resulting in further breakdowns.
Note that different from the PMA domain walls \cite{Laurson20191,tilting1, tilting2, tilting3} and HtH/TtT domain walls \cite{Yuan2014}, there is no global, directional in-plane tilting of the domain wall centerline  {(domain wall centerline is the contour line of $m_y=0$, i.e. the domain wall center)}. Of course, transient, local tilting near the vortices is ubiquitous, as shown in Fig. \ref{fig3}(a) and (b). This is because in the side-by-side geometry, there are no magnetic charges at two  {ends of the domain wall} like those in PMA and HtH/TtT domain walls, so the domain wall width $\Delta$ is almost constant along $y$ direction. For the snapshots shown in Fig. \ref{fig3}(a)(b), the difference between the smallest and largest $\Delta$ is less than $3\%$.

\section{INFLUENCE OF Dzyaloshinskii-Moriya interaction}
\begin{figure}[ht]
	\includegraphics[width=0.45\textwidth]{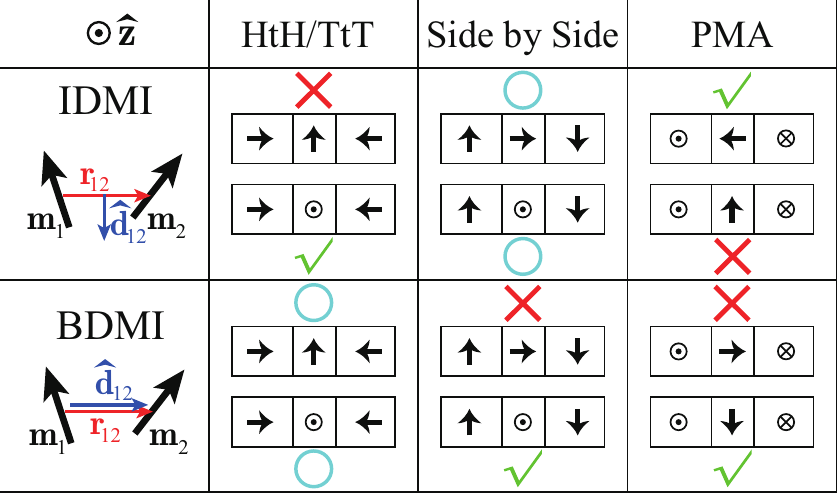}
	\caption{Schematic diagrams of interfacial and bulk DMI and summary of energetically preferred static domain wall configurations.
The check mark (cross mark) means the configuration is preferable (not preferable). The circle means that the DMI has no
influence on the static domain wall configuration. }
	\label{fig4}
\end{figure}

As an antisymmetric exchange interaction, DMI has been demonstrated to have a significant influence on HtH/TtT \cite{Zhuo2016}
and PMA \cite{Thiaville2012,Emori2013} domain wall dynamics. There are two most widely studied types of DMI, i.e. the interfacial DMI
(IDMI) and the bulk DMI (BDMI). The interfacial DMI exists in inversion symmetry breaking systems. The DMI vector direction
 $\hat{\mathbf{d}}_{12}$ from
spin 1 to spin 2 is parallel to $\mathbf{r}_{12}\times \hat{\mathbf{z}}$, where $\mathbf{r}_{12}$ is the spatial vector from 1 to 2 and
$\hat{\mathbf{z}}$ is the inversion symmetry breaking direction \cite{Thiaville2012}. The bulk DMI exists in noncentrosymmetric systems.
$\hat{\mathbf{d}}_{12}$ is parallel to $\mathbf{r}_{12}$ \cite{Nagaosa2013}. The DMI vector directions are schematically illustrated in
Fig. \ref{fig4}. For static domain wall configurations, it is enough to use the simplest three-spin model to decide the energetically preferred configuration in quasi-1D. Figure \ref{fig4} summarizes the
influence of the two types of DMI on different types of domain walls. The side-by-side walls are different from the HtH/TtT walls and PMA walls. The IDMI does not break the degeneracy of N\'{e}el type (domain wall center in-plane) and the Bloch type (domain wall center out-of-plane). The BDMI prefers the Bloch-type, which is competing with the shape anisotropy. In the continuous model, the expressions of energy density of IDMI and BDMI are, respectively,
\begin{gather}
E_\mathbf{IDMI}=D_i\left[m_z\nabla\cdot\mathbf{m}-(\mathbf{m}\cdot\nabla)m_z\right],\\
E_\mathbf{BDMI}=D_b \mathbf{m}\cdot(\nabla\times \mathbf{m}),
\end{gather}
where $D_i$ and $D_b$ are the IDMI and BDMI strength in units of $\text{J/m}^2$. Applying the quasi-1D $X-\Phi$ collective coordinate
model \eqref{ccm}, we find the total energy $\mathcal{E}$ of the side-by-side domain wall in the presence of IDMI or BDMI
(see Appendix),
\begin{gather}
\mathcal{E}_\mathbf{IDMI}=4dw\sqrt{A\left(K_y+K_z\cos^2{\Phi}\right)},
\label{IDMI1D}\\
\mathcal{E}_\mathbf{BDMI}=dw\left[4\sqrt{A\left(K_y+K_z\cos^2{\Phi}\right)}-\pi D_b \cos\Phi\right].
\end{gather}

\begin{figure}[ht]
	\includegraphics[width=0.45\textwidth]{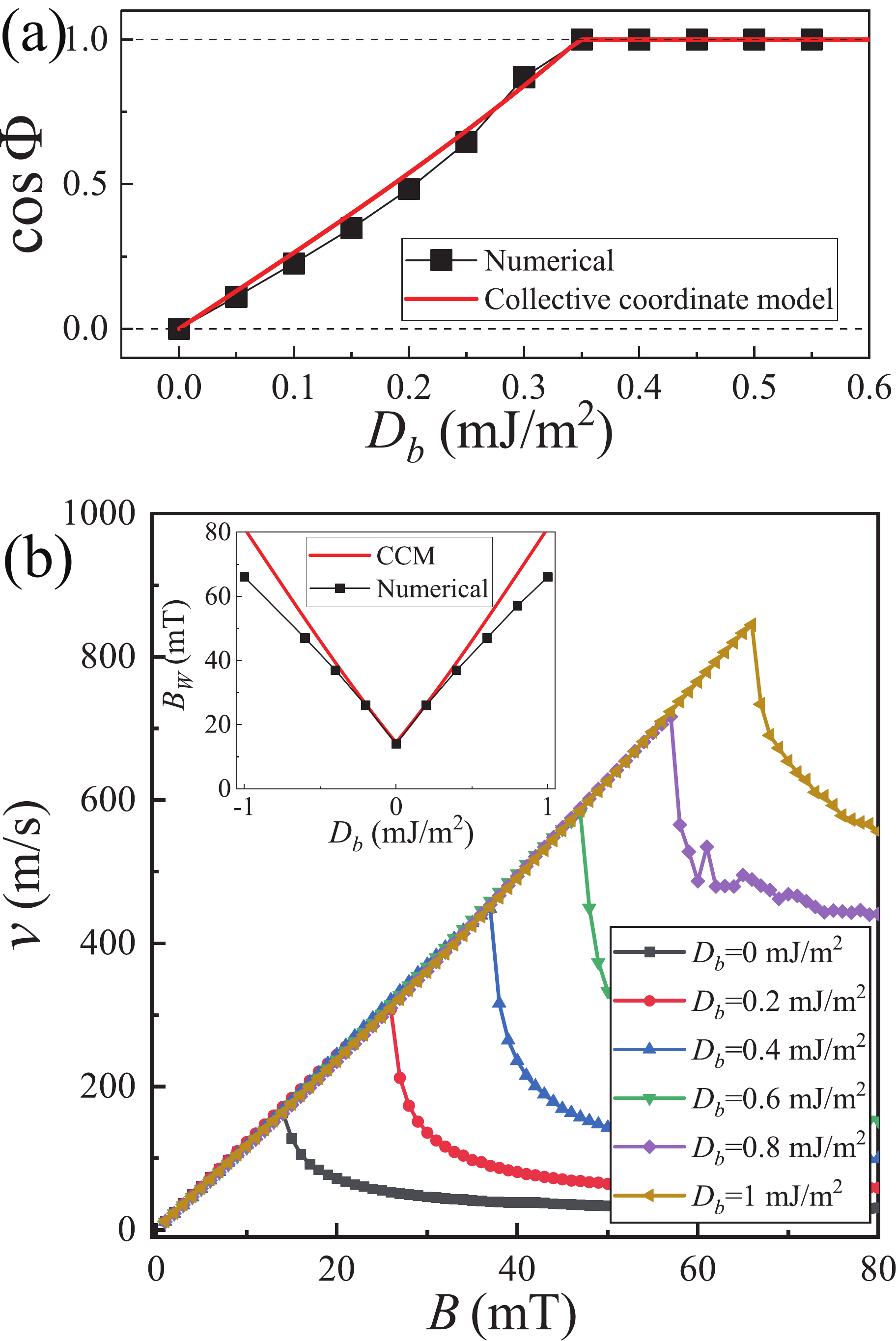}
	\caption{(a) Influence of BDMI strength $D_b$ on domain wall azimuthal angle $\Phi$. The symbols are numerical data and the solid line is the collective coordinate model result. (b) The simulation results of field-driven domain wall velocity for $w=48$ nm and different $D_b$. The inset shows the Walker breakdown field $B_W$ versus $D_b$. The symbols are numerical data and the solid line is the collective
coordinate model result.}
	\label{fig5}
\end{figure}

We first focus on the influence of BDMI. For the static domain wall configuration, minimizing $\mathcal{E}_\mathbf{BDMI}$ with respect to $\Phi$, we find
\begin{equation}
\cos \Phi=\left\{
\begin{array}{ll}
\pi D_b \sqrt{\frac{K_y}{K_z(16AK_z-\pi^2D_b^2)}}& |D_b|<D_c\\
\mathrm{sign}(D_b) & |D_b|\geq D_c
\end{array}\right.
\label{CCMBDMI}
\end{equation}
where $D_c=\frac{4K_z}{\pi}\sqrt{\frac{A}{K_y+K_z}}$. For $w=48$ nm, with $K_y=2.09\times10^5$ J/m$^3$ and $K_z=4.337\times 10^4$ J/m$^3$ obtained in the previous section, we have $D_c= 0.348$ $\text{mJ/m}^2$. From $D_b=0$ to $D_b=D_c$, the static domain wall gradually rotates from N\'{e}el-type ($\Phi=\pm\pi/2$) to Bloch-type ($\Phi=0$ for positive $D_b$ and $\Phi=\pi$ for negative $D_b$). Figure \ref{fig5}(a) shows the quasi-1D result Eq. \eqref{CCMBDMI} together with the numerical results, showing a reasonably good agreement.

Then we investigate the field-driven dynamics of side-by-side domain walls in the presence of BDMI. We have discussed that the BDMI tends to lock the domain wall in Bloch type [$D_b>0$ ($<0$) for domain wall center pointing to $+z$ ($-z$)].
Thus, when an external field is applied, the domain wall rotation is suppressed so that the Walker breakdown is postponed.
Figure \ref{fig5}(b) shows how the average domain wall velocity changes with external field for the 48 nm strip.
Typical Walker breakdown behaviors are observed.
The breakdown field $B_W$ increases with $D_b$. The breakdown fields $B_w$ for different $D_b$ are plotted in the inset. The black symbols
are the numerical results extracted from the main figure, and the red line is the result of 1D CCM (see Appendix). The numerical results qualitatively compare well with the 1D CCM model, but the $B_W$ values are smaller, mainly due to the
2D nature and the complicated demagnetization field in the numerical model. Both the CCM model and numerical data show that the domain
wall velocity is symmetric for positive and negative $D_b$.

\begin{figure}[ht]
	\includegraphics[width=0.45\textwidth]{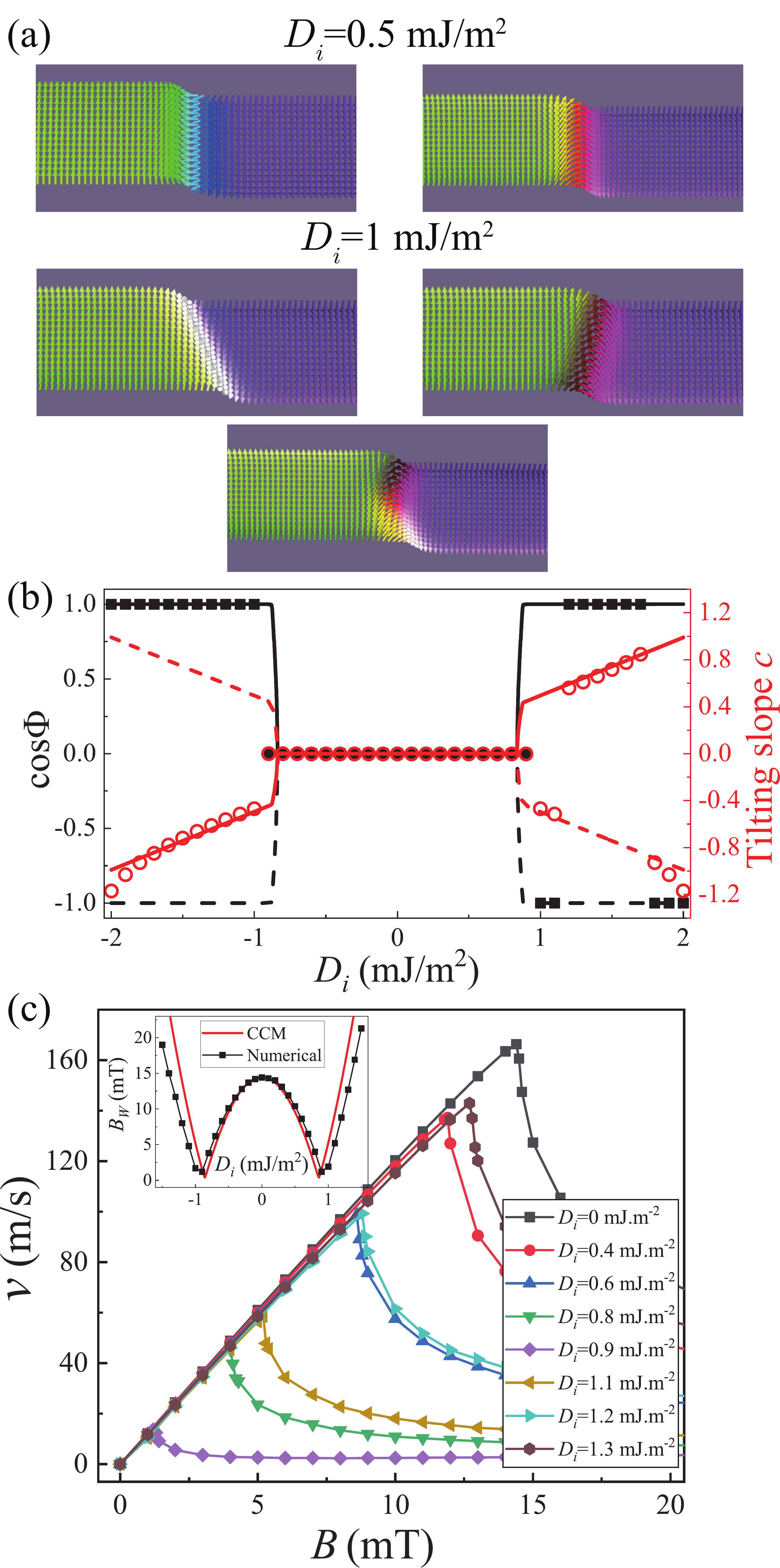}
	\caption{(a) Static domain wall configuration for $w=48$ nm and $D_i=0.5$ (first row) and 1 $\text{mJ/m}^2$ (second row) numerically relaxed from different random initial states. (b) $\cos\Phi$ of domain wall plane (left axis) and tilting slope $c$ of domain wall centerline (right axis). The symbols are numerical data and the lines are 2D CCM results. The solid lines and dashed lines are two possible combinations of $\cos\Phi$ and $c$. (c) Field-driven domain wall velocity for different $D_i$. Inset: Walker breakdown field
$B_W$ versus $D_i$. Symbols are numerical data and the solid line comes from the 2D CCM.}
	\label{fig6}
\end{figure}

We now consider IDMI. According to Eq. \eqref{IDMI1D}, the IDMI does not affect the static domain wall configuration in 1D model.
For a 2D strip of $w=48$ nm, this is still true for weak IDMI such as $D_i=0.5$ $\text{mJ/m}^2$. However,
the numerical relaxation shows that the static domain wall centerline is tilted in-plane, and the magnetization is tilted out-of-plane, as shown in Fig. \ref{fig6}(a) for $D_i=1$ $\text{mJ/m}^2$. The tilting direction of the domain wall centerline is correlated with the tilting direction of the magnetization. For $D_i>0$, when the domain wall centerline lies in the first and third (second and fourth) quadrants, the domain wall magnetization is tilted to $-z$ ($+z$). The probability of the two tilting directions is
the same for different random initial states. For some initial states, it is also possible to have more complicated domain wall
texture, such as that shown in the third panel of Fig. \ref{fig6}(a). Different segments of the domain wall tilt to different directions.
Such DMI-induced tilting has been observed in PMA domain walls \cite{tilting1, tilting2, tilting3}. To explain the tilting, we
have to introduce a 2D collective coordinate model, allowing $X$ to be dependent on $y$, $X=X(y,t)$. We assume the tilting is linear
so that $\frac{dX}{dt}=c$ is constant. The static domain wall energy is,
\begin{multline}
\mathcal{E}_\mathbf{IDMI}\\=dw\left[2(2+c^2)\sqrt{A\left(K_y+K_z\cos^2{\Phi}\right)}+\pi c D_i \cos\Phi\right].
\label{dwenergyi}
\end{multline}
The first term in the bracket is related to the balance between exchange energy and anisotropy energy. The larger the tilting, the longer the domain wall, so the first term prefers smaller $c$. The second term is related to the IDMI. Only when this term is negative, the tilted domain wall can be possibly preferred. $\mathcal{E}_\mathbf{IDMI}$ can be minimized with respect to $c$ and $\Phi$. For $D_i>0$, there are two degenerated minimal points, corresponding to $c>0$, $\Phi>\pi/2$ and $c<0$, $\Phi<\pi/2$, respectively, which is consistent with the numerical results. Figure \ref{fig6}(b) plots the domain wall azimuthal angle $\cos\Phi$ (left axis) and the tilting slope $c$ (right axis), showing the comparison between numerical data and the CCM model. The solid lines and the dashed lines represent the two
ways of tilting with the same energy. The numerical data almost fall on either the solid lines or the dashed lines.
When applying an external field, the side-by-side domain wall dynamics shows more interesting behaviors in the presence of IDMI. Figure
\ref{fig6}(c) exhibits the domain wall velocity versus applied field for different $D_i$. The Walker breakdown field $B_W$ first decreases, then increases with $D_i$, which is different from the HtH/TtT and PMA domain walls. This phenomena can also be understood using the 2D CCM. For small $D_i$, the straight N\'{e}el domain wall is still the ground state, and the tilted domain wall only has slightly higher energy. Their energy difference plays a role of a low energy barrier. When $D_i$ increases, the energy of tilted domain wall becomes lower, so that the domain wall is easier to flip between the two N\'{e}el configurations, leading to a lower $B_W$. After a certain value of $D_i$, the tilted domain wall becomes the ground state, and the energy barrier becomes higher when further increasing $D_i$. Thus, the flipping of domain wall center becomes more difficult and $B_W$ increases. The inset of Fig. \ref{fig6}(c) shows $B_W$ from numerical data, and the CCM result is plotted in solid line for comparison. The simulation and CCM agree well with each other.
However, the CCM indicates a symmetrical $B_W$ in positive and negative sides of $D_i$, but the numerical results are asymmetric. $B_W$
for $-|D_i|$ is always slightly smaller than that for $+|D_i|$. This qualitative discrepancy is mainly due to the IDMI-induced tilting
of magnetization at the two side edges of the strip. We recall the finding in Section \ref{MTWB} that for $B>0$, the inhomogeneous flipping of domain wall center always starts at the bottom edge. This is also true in the presence of DMI. For positive $D_i$, at the bottom edge, the magnetization in the left (right) domain is tilted towards $-z$ ($+z$) to minimize the IDMI energy [this can be observed
in Fig. \ref{fig6}(a)]. This tilting is clockwise with respect to $+y$ direction, which is opposite to the counterclockwise torque induced
by $B$. Thus, the bottom edge is relatively robust so that larger field is required to flip the magnetization inside the domain wall.
On the contrary, for negative $D_i$, the tilting is along the same direction as the torque of $B$. So the breakdown field is smaller.
Since the tilting at the edge is small, this difference in $B_W$ is subtle. For larger $|D_i|$, the difference becomes more significant.

\begin{figure}[ht]
	\includegraphics[width=0.45\textwidth]{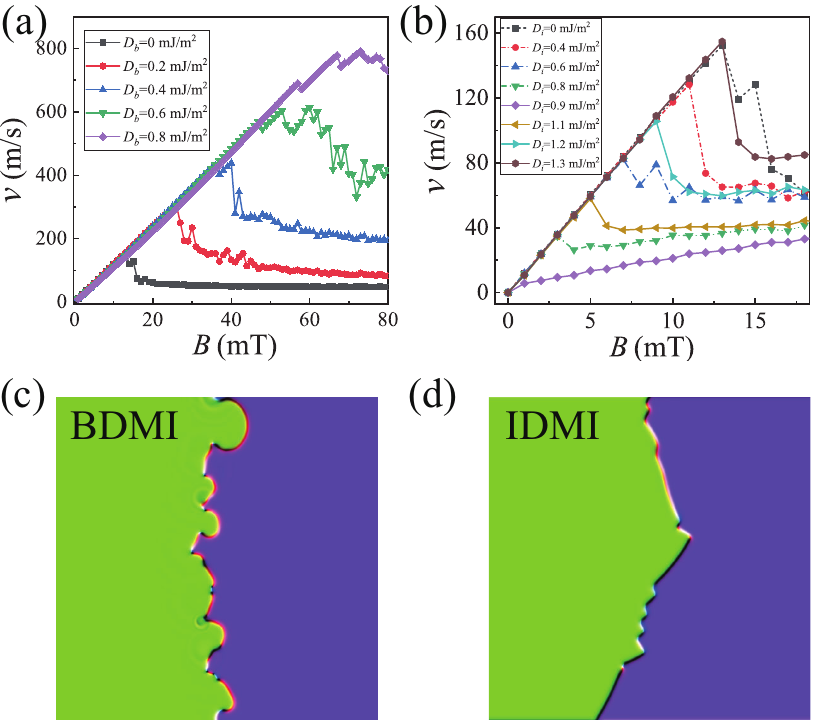}
	\caption{Domain wall velocity $v$ versus external field $B$ for 1536 nm strip in the presence of different (a) BDMI, (b) IDMI.
Typical domain wall texture after Walker breakdown for (c) BDMI, $D_b=0.8$ mJ/m$^2$ and $B=150$ mT, (d) IDMI, $D_i=1.3$ mJ/m$^2$ and $B=15.2$ mT. }
	\label{fig7}
\end{figure}

We also perform simulations for wider strips in the presence of BDMI and IDMI. Figure \ref{fig7} shows the field-driven domain wall
velocity for $w=1536$ nm. Before the Walker breakdown, the dynamics does not differ too much from the 48 nm strip, except
the breakdown field becomes smaller.  {After the Walker breakdown, the dynamics affected by BDMI and IDMI are distinct.}

 {In the presence of BDMI, the energy degeneracy of two chiralities is broken, and the system can be mapped to the well-studied
PMA domain walls by a $\pi/2$ rotation of $\mathbf{m}$ around $x$. Thus, we also observe soliton-like domain wall motion similar to that of PMA domain wall described in Ref. \cite{Ono2016}. A typical domain wall texture is shown in Fig. \ref{fig7}(c) for $w=1536$ nm, $D_b=0.8$ mJ/m$^2$ and $B=150$ mT. The snapshot is taken at $t=1$ ns. A tortuous domain wall centerline with many vortices can be observed. The vortex generation and annihilation are similar to the Bloch line in PMA domain wall \cite{Ono2016}. The suppression of Walker breakdown is also present for $D_b$ larger than 1 mJ/m$^2$. More details are discussed in Appendix \ref{AC}. }

 {The IDMI has totally different affect on the domain wall dynamics. Figure \ref{fig7}(d) shows a typical domain wall texture for $w=1536$ nm, $D_b=1.3$ mJ/m$^2$ and $B=15.2$ mT. Different from the curled, tortuous domain wall in the BDMI case, the domain wall centerline is zigzag with each segment straight and the magnetization is out-of-plane for each segment. The tilting direction follows the same rule as the static case discussed above. Compared to the fast ($\sim 0.1$ ns) vortex dynamics in the BDMI case, the zigzag domain wall deforms slowly. New zigzags appear at the edges and gradually annihilate at the middle. Since the dynamics in the presence of DMI is quite complicated, we will study it in a statistical way in future.}

\section{Discussion}
In the calculations above, we use a large anisotropy $K_u$ so that the domain wall is thin ($\sim 7$ nm) to avoid the influence of finite strip length as much as possible. The validity of continuous model can be demonstrated by comparing the numerical domain wall profile with the Walker solution [the inset of Fig. \ref{fig1}]. Furthermore, the threshold domain wall width between continuous domain wall and abrupt domain wall is $\Delta=\sqrt{3}a/2$ where $a$ is the lattice constant (the mesh size in our case) \cite{JMMM1994,Yanpeng2012}. Our domain wall width is above this threshold. For wider domain walls, our results are still qualitatively correct.

We have explained the observed dynamics in the energy point of view. The Zeeman energy is dissipated via Gilbert damping leading to propagation of domain wall, and the transit vortex generation processes temporarily store the Zeeman energy leading to the change in domain wall speed. We should note that the energy argument can only give an overall understanding, but cannot give the detailed dynamics. For weak field below the Walker breakdown, the Zeeman energy can be solely dissipated through Gilbert damping. So that the magnetic texture is able to keep unchanged, resulting in the rigid-body motion. When the field is closed to or slightly larger than the Walker breakdown, spin wave emission may occur to dissipate more energy \cite{SW2010,Field4}. For larger field, domain wall starts to rotate (in thin strips) or nucleate vortices (in wider strips). The Zeeman energy is periodically stored and released by the domain wall, as we have discussed above. For even larger field (larger than the effective field of the easy-axis anisotropy), the domain opposite to the field becomes unstable and more complicated textures emerge, such as another pair of domain walls. In different situations the Zeeman energy dissipates and converts in different ways. Analyses on the LLG equation are still necessary to know the specific dynamics.

Materials with in-plane uniaxial anisotropy are necessary to experimentally observe and investigate the side-by-side domain walls. The anisotropy $K_u$ should overcome the in-plane shape anisotropy $(N_y-N_x)\mu_0M_s^2$. It has been observed that cobalt can possess such $K_u$ \citep{Cobalt2000}. The growing condition or external strain can also induce an in-plane uniaxial anisotropy \cite{Py2020,Gilbert2017}. The IDMI may be present in such materials by designing inversion-symmetry structures.
It is also possible to induce in-plane uniaxial anisotropy by strain in BDMI materials \cite{Shibata2015}. Of course, we have studied an ideal theoretical model here. Finite temperature, geometrical defects like edge roughness and surface roughness, and material defects (including inhomogeneity of material parameters) would exist in reality. They will be the topics of further theoretical studies. In the Appendix, we briefly discuss the influence of inhomogeneous $K_u$, which is supposed to be ubiquitous in imperfect materials.

\section{Summary}
We investigate the static and dynamic properties of side-by-side domain walls.
Although the observed side-to-side domain wall dynamics has many similarities to the HtH/TtT and PMA domain walls, there still exist important differences due to the different geometries.
In the absence of DMI, the domain wall is in N\'{e}el configuration due to the shape anisotropy,
and the domain wall width $\Delta$ can be estimated self-consistently using the shape anisotropy of a prism of dimensions
$(\Delta, w,d)$. The field-driven domain wall dynamics in thin strips can still be described by 1D collective coordinate model.
In wide strips, complicated multistep breakdown behavior occurs via generation, propagation and annihilation of (anti)vortices. Due to the absence of magnetic charges at the ends of the domain wall, the side-by-side domain wall width is more homogeneous than the other two kinds of domain walls, and there is no directional tilting.
In the presence of BDMI, the Walker breakdown field increases with the BDMI so that the fast rigid-body domain wall motion is enhanced.
The simulation results for thin strips can be well reproduced by the 1D collective coordinate model.
In the presence of IDMI, domain wall tilting occurs at
strong IDMI. The Walker breakdown field first decreases and then increases with the IDMI strength.
The non-monotonic dependence of breakdown field on IDMI as well as the domain wall tilting can be explained by 2D collective coordinate model. Furthermore, the breakdown field shows subtle asymmetry in positive and negative IDMI, mainly due to the IDMI-induced magnetization
tilting at the strip edges.
For wider strips, in the presence of BDMI, the domain wall is tortuous with plenty of vortex generation and annihilation. Soliton-like dynamics similar to the PMA domain wall case is observed. In the presence of IDMI, the domain wall becomes zigzag. Our results provide more comprehensive understandings on the properties of domain walls.
	
\begin{acknowledgments}
	This work is supported by the Fundamental Research Funds for the Central Universities. X. S. W. acknowledges the support from the Natural Science Foundation of China (NSFC) (Grant No. 11804045 and No. 12174093). F. X. L. acknowledges the support from the Natural Science Foundation of China (NSFC) (Grant No. 11905054).

\end{acknowledgments}

\appendix
\section{\label{AA}Collective Coordinate Model}
The collective coordinate model has been comprehensively studied in previous research \cite{Thiaville2006,Shibata2011,Thiaville2012,tilting1}.
Here we emphasize what are different in side-by-side domain walls. As depicted in Fig. \ref{fig1}, we define the spherical coordinate
with respect to $y$ axis for convenience, so that $m_y=\cos \theta$, $m_z=\sin\theta \cos\phi$, $m_x=\sin\theta \sin\phi$.
With the effective shape anisotropy, the total energy of the system is
\begin{multline}
\mathcal{E}=\int \bigg[A\left|\nabla \mathbf{m}\right|^2+K_z (\mathbf{m}\cdot \hat{\mathbf{z}})^2-K_y (\mathbf{m}\cdot \hat{\mathbf{y}})^2
\\-M_s \mathbf{m}\cdot \mathbf{B}\bigg]dV
+\int E_\text{IDMI(BDMI)} dV
\end{multline}
In spherical coordinates, the DMI energy density should be written as,
\begin{gather}
E_\text{IDMI}=D_i\left(\sin^2\theta\frac{\partial \phi}{\partial x}-\cos\phi\frac{\partial \theta}{\partial y}+\frac{\sin2\theta}{2}\sin\phi\frac{\partial \phi}{\partial y}\right),\\
E_\text{BDMI}=D_b\left(-\sin^2\theta\frac{\partial \phi}{\partial y}-\cos\phi\frac{\partial \theta}{\partial x}+\frac{\sin2\theta}{2}\sin\phi\frac{\partial \phi}{\partial x}\right)
\end{gather}
We introduce a Lagrangian representation \cite{Shibata2011} with kinetic term
\begin{equation}
\mathcal{L}_T=-\frac{M_s}{\gamma}\int \cos\theta \frac{\partial \phi}{\partial t} dV
\end{equation}
and dissipation term
\begin{equation}
\mathcal{W}=\frac{M_s}{2\gamma}\int\left[ \left(\frac{\partial \theta }{\partial t}\right)^2+\sin^2 \theta \left(\frac{\partial \phi}{\partial t}\right)^2 \right]dV.
\end{equation}
The Lagrangian is $\mathcal{L}=\mathcal{L}_T-\mathcal{E}$. The Euler-Lagrangian equations
\begin{equation}
\frac{d}{dt}\frac{\partial L}{\partial(\partial_t q)}-\frac{\partial L}{\partial q}=-\frac{\partial W}{\partial (\partial_t q)},
\end{equation}
reproduce the LLG equation \eqref{LLGeq}, where $L,W$ are integrands of $\mathcal{L},\mathcal{W}$, and $q$ represents $\theta$ or $\phi$.

The collective coordinate model (CCM) assumes the following planar Walker domain wall profile,
\begin{equation}
\theta(x,y,t)= 2\arctan (e^{\frac{x-X(t)}{\Delta}}),\quad \phi(x,y,t)=\Phi(t),
\label{ccm}
\end{equation}
with domain wall width
\begin{equation}
\Delta=\sqrt{\frac{A}{K_y+K_z\cos^2\phi}}.
\end{equation}
We further assume $K_z\ll K_y$ so that the deformation of the domain wall is insignificant and $d\Delta/dt$ can be ignored.
The 1D CCM model further assumes the magnetization is uniform in $y$ and $z$ directions.
In the absence of DMI, the CCM for side-by-side domain walls is the same as the HtH/TtT or PMA walls, which has been well studied
\cite{Thiaville2006,Shibata2011}. In the presence of BDMI, the 1D CCM Lagrangian and dissipation function can be obtained by substituting Eq. \eqref{ccm} into $\mathcal{L}$ and $\mathcal{W}$. For the non-convergent integral $\int \cos\theta dx$, we consider the principal value $\lim_{L\rightarrow\infty}(\int_{-L}^{L}\cos\theta dx)=2X$. Thus, we have
\begin{gather}
\mathcal{L}=dw\left(-2X\frac{M_s}{\gamma}\frac{d\Phi}{d t}+2M_sBX\right)-\mathcal{E}_\text{BDMI},\\
\mathcal{W}=\alpha dw\frac{M_s}{\gamma}\left[\frac{1}{\delta}\left(\frac{dX}{dt}\right)^2+\Delta \left(\frac{d\Phi}{dt}\right)^2\right].
\end{gather}
The equations of motion can be obtained by the Euler-Lagrangian equations with respect to $q=X$ and $q=\Phi$,
\begin{equation}
\frac{d}{dt}\frac{\partial \mathcal{L}}{\partial(\partial_t q)}-\frac{\partial \mathcal{L}}{\partial q}=-\frac{\partial \mathcal{W}}{\partial (\partial_t q)}.
\end{equation}
We have
\begin{gather}
\frac{d\Phi}{dt}+\alpha\frac{1}{\Delta}\frac{dX}{dt}=\gamma B,\\
\frac{M_s}{\gamma}\left(\alpha\frac{d\Phi}{dt}-\frac{1}{\Delta}\frac{dX}{dt}\right)=K_z\sin2\Phi-\frac{\pi D}{2\Delta}\sin\Phi.
\end{gather}
For the rigid body motion, $\partial\Phi/\partial t=0$, we have the relation
\begin{equation}
B=\frac{\alpha}{M_s}\left(K_z\sin2\Phi-\frac{\pi D}{2\Delta}\sin\Phi\right)
\end{equation}
Thus, the Walker breakdown field $B_W$ is the maximal value of the right-hand side, above which the rigid body motion is invalid.
This value can be calculated straightforwardly after some tedious mathematics. The results are plotted in the inset of Fig. \ref{fig5}(b).

In the presence of IDMI, we introduce 2D CCM model by allowing $X=X(y,t)$. For simplicity, we assume $\partial X/\partial y=c$ is a constant. The CCM gives the domain wall energy Eq. \eqref{dwenergyi}, and the corresponding Lagrangian
\begin{equation}
\mathcal{L}=dw\left(-2X_0\frac{M_s}{\gamma}\frac{d\Phi}{d t}+2M_sBX_0\right)-\mathcal{E}_\text{IDMI},
\end{equation}
where $X_0$ is the domain wall center position at $y=0$ (the centerline of the strip).
Notice that besides $X_0$ and $\Phi$, $c$ is also a variable and has its own Euler-Lagrangian equation $\partial\mathcal{L}/\partial c=0$,
giving
\begin{equation}
c=\frac{-\pi D_i \cos\Phi}{4\sqrt{A(K_y+K_z\cos^2\Phi)}}.
\label{crela}
\end{equation}
The second equation of motion is modified as,
\begin{multline}
\frac{M_s}{\gamma}\left(\alpha\frac{d\Phi}{dt}-\frac{1}{\Delta}\frac{dX}{dt}\right)\\
=(1+c^2)K_z\sin2\Phi+\frac{\pi D c}{2\Delta}\sin\Phi.
\end{multline}
We can similarly obtain the relation for rigid body motion,
\begin{equation}
B=\frac{\alpha}{M_s}\left[(1+c^2)K_z\sin2\Phi+\frac{\pi D c}{2\Delta}\sin\Phi\right]
\end{equation}
Substituting Eq. \eqref{crela} into the above relation, and taking the maximal value,
we can find the breakdown field $B_W$ for IDMI.

\section{\label{AB}Influence of Inhomogeneous Anisotropy}
\begin{figure}[ht]
	\includegraphics[width=0.45\textwidth]{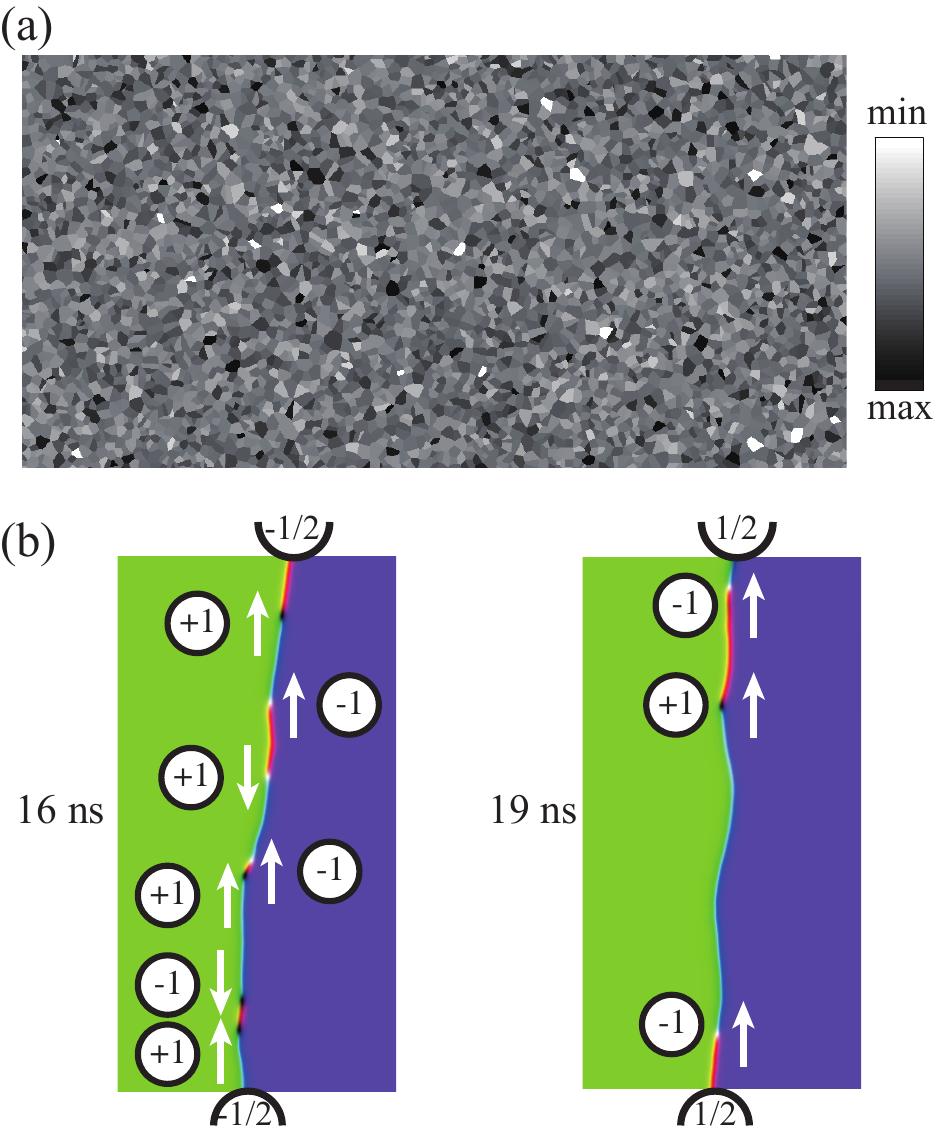}
	\caption{(a) A typical grain tessellation and the corresponding anisotropy distribution (the grayscale colorbar).
(b) Snapshots of domain wall dynamics at 16 ns and 19 ns under $B=13.8$ mT.}
	\label{fig8}
\end{figure}

Practically, the material may be inhomogeneous so that the material parameters are position-dependent. Here we briefly discuss the influence of inhomogeneous anisotropy $K_u$. We divide the 1536 nm wide strip into 256 grains using Voronoi tessellation \cite{mumax}.
Each grain is assigned a random number $r$ following the standard normal distribution, and the anisotropy of the grain is $(1+0.1r)K_u$.
A typical grain tessellation and the corresponding anisotropy distribution are shown in Fig. \ref{fig8}(a) (the brightness of the color encodes the anisotropy strength). Due to the inhomogeneous anisotropy, the generation of vortices becomes more complicated.
Figure \ref{fig8}(b) shows the snapshots of domain wall dynamics at 16 ns and 19 ns under $B=13.8$ mT. Different from Fig. \ref{fig3}(a)(b), there are multiple two-vortex processes and vortices with different polarities coexisting. The total winding number is still 0. The direction of transverse motion of a vortex depends on the product of vorticity and polarity, as labelled in the figure. The average domain wall speed is smaller than the homogeneous sample. More domain wall dynamics in the presence of different kinds of defects and randomness will be studied in future.

\section{\label{AC}Domain wall dynamics in the presence of DMI}

\begin{figure}[ht]
	\includegraphics[width=0.45\textwidth]{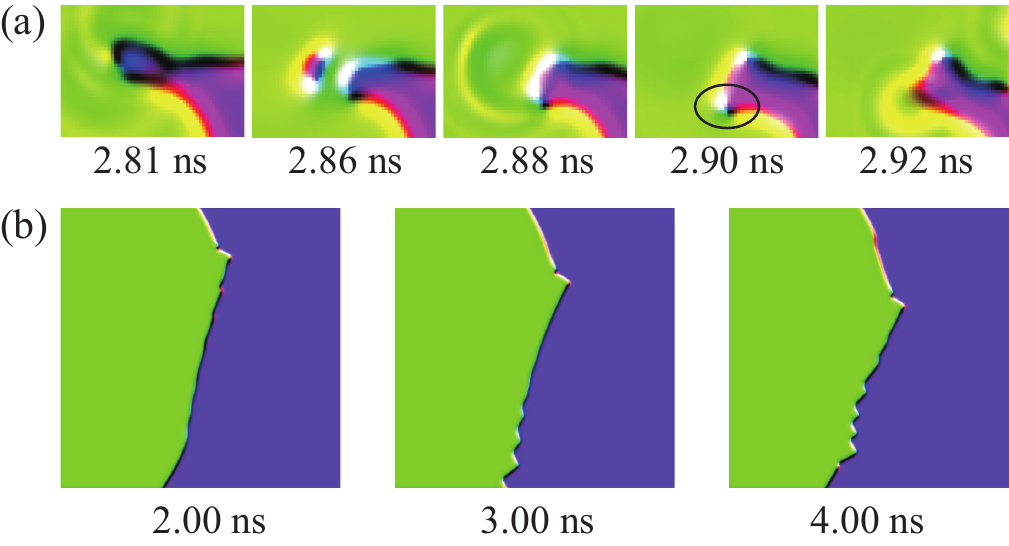}
	\caption{(a) Close-up view of a segment of domain wall for $w=1536$ nm, $D_b=0.8$ mJ/m$^2$ and $B=150$ mT. (b)
Domain wall for $w=1536$ nm, $D_i=1.3$ mJ/m$^2$ and $B=15.2$ mT.  }
	\label{fig9}
\end{figure}

 {We first discuss the BDMI. We observe similar soliton-like dynamics as Ref. \cite{Ono2016}. Figure \ref{fig9}(a) shows close-up snapshots of a segment of domain wall. Due to the BDMI, the energy of vortices of opposite chirality becomes different, and their transverse speed is also different. Here we denote the vorticity (1D winding number) by $V$, the polarity by $P$, and the skyrmion number (2D winding number) by $Q$. Then for a local vortex, $Q=\frac{1}{2}VP$ \cite{Oleg2005,WXS2021}. At 2.90 ns, a $V=-1$, $P=-1$ antivortex collides with a $V=+1$, $P=+1$
vortex (indicated by the black circle). Then they annihilate and spin wave is emitted (the spin wave ripple can be seen in the 2.92 ns snapshot). This procedure is different from that in the absence of DMI. In the absence of DMI, the collision can only happen between a vortex ($V=+1$ and an antivortex ($V=-1$) of same $P$ who move in opposite directions. Both total $V$ and total $Q$ are conserved to be 0, and the annihilation is smooth with negligible spin wave emission. However, in the presence of BDMI, the $V=-1$, $P=-1$ antivortex and the $V=+1$, $P=+1$
vortex have same $Q=+\frac{1}{2}$ and move in the same direction. $V$ is conserved but $Q$ is not conserved during the annihilation. Thus, significant spin wave emission can be observed.
We also identify another kind of spin wave emission mechanism (we note that this can also be observed in the PMA case). See the 2.81 ns and 2.86 ns snapshots. A bubble of $Q=+1$ detaches from the domain wall due to the severe distortion. Then the bubble blasts, accompanied with significant spin wave emission (see the ripples in the 2.88 ns snapshot).}

 {Then we discuss the IDMI. In the presence of IDMI, the domain wall becomes zigzag, and the magnetization of the domain wall center is mainly out-of-plane except the transition regions between adjacent segments. The tilting direction of each segment follows the rule discussed in the main text for static domain walls. Zigzags gradually emergent at edge and inside the long segments, and annihilate at the middle. It can also be seen that compared to the fast vortex dynamics in the BDMI case ($\sim 0.1$ ns), the zigzags evolution is much slower. The zigzags are not local textures like vortices, but the transition regions can also be treated as vortices. We will study the dynamics of the local textures (solitons) in statistical point of view in future.}

\end{document}